\documentclass[cits]{PoS} 
\title{Charmed-Meson Decay Constants from Improved QCD Sum Rules}
\ShortTitle{Charmed-Meson Decay Constants from Improved QCD Sum
Rules}\author{\speaker{Wolfgang Lucha}\\Institute for High Energy
Physics, Austrian Academy of Sciences, Nikolsdorfergasse 18,
A-1050 Vienna, Austria\\E-mail: \email{Wolfgang.Lucha@oeaw.ac.at}}
\author{Dmitri Melikhov\\Institute for High Energy Physics,
Austrian Academy of Sciences, Nikolsdorfergasse 18, A-1050 Vienna,
Austria,\\Faculty of Physics, University of Vienna, Boltzmanngasse
5, A-1090 Vienna, Austria, and\\D.~V.~Skobeltsyn Institute of
Nuclear Physics, Moscow State University, 119991, Moscow,
Russia\\E-mail: \email{dmitri\_melikhov@gmx.de}}\author{Silvano
Simula\\INFN, Sezione di Roma Tre, Via della Vasca Navale 84,
I-00146 Roma, Italy\\E-mail: \email{simula@roma3.infn.it}}

\abstract{The decay constants $f_{D_{(s)}}$ of the charmed heavy
pseudoscalar mesons $D$ and $D_s$ are revisited within a recently
developed novel approach to dispersive QCD sum rules which relies
on an unprejudiced implementation of quark--hadron duality. The
proposed modification of the conventional sum-rule techniques is
assessed by applying our prescriptions to quantum mechanics, where
exact solutions may be easily obtained by simply solving the
Schrödinger equation. The very striking similarity of the
extraction procedures of bound-state parameters in potential
models and in QCD gives~us great confidence in the reliability of
our improvements of the sum-rule formalism and their applicability
to hadron phenomenology. The implications of one's chosen
definition of the heavy-quark masses are scrutinized and the
$\overline{\rm MS}$ quark-mass scheme is identified as the optimal
choice for our purposes. Our ideas turn out to reconcile QCD
sum-rule predictions for the charmed-meson decay constants, which
before tended to be markedly too low, with the findings of both
lattice QCD and experiment.}

\FullConference{The 2011 Europhysics Conference on High Energy
Physics -- HEP 2011,\\July 21--27, 2011\\Grenoble, Rh\^one-Alpes,
France}

\begin{document}
\section{Introduction: Improvement of QCD Sum Rules as Incentive
and Main Aim \cite{LMS:SUE,LMS:FF,LMS:ECT,LMS:QCD,LMS:HMDC}}Unlike
lattice gauge theories, QCD sum rules allow for fully {\em
analytic\/} investigations of hadrons: {\em QCD sum rules\/}
relate hadronic features to QCD by evaluation of matrix elements
of suitably chosen operators on both hadron and QCD level and by
assuming {\em quark--hadron duality\/}: above an {\em effective
threshold\/} the perturbative QCD contribution cancels the one of
hadronic excitations and continuum. Regarding the threshold as
function of a parameter entering upon Borel transformation, we
improve this concept by straightforward simple techniques
\cite{LMS:SUE,LMS:FF,LMS:ECT} enabling us to estimate intrinsic
errors too.

\section{Nonrelativistic Potential Models in Quantum Mechanics as
Test Area of our Ideas}The lack of a clear route to all their
errors is a {\em not unsurmountable\/} weakness of QCD sum rules:
Modelling the strong interactions by the funnel potential (or the
like) describing heavy-quark bound states \cite{LSG:QBS}, we
demonstrate \cite{LMS:QCD} the capability of our step by
application to quantum mechanics. There the {\em exact\/}
bound-state features can be derived by numerical solution
\cite{Lucha98} of the Schr\"odinger equation. Requiring the
duality-truncated QCD member of the sum rule to counterbalance
{\em exactly\/} its hadronic counterpart, that is, the lowest
hadronic term, we prove that the induced {\em exact\/} threshold
depends not only on the external momenta involved but also on the
Borel parameter and the underlying~operator.

\section{Decay Constants $f_{D}$ and $f_{D_s}$ of the Charmed
Pseudoscalar Mesons $D$ and $D_s$ \cite{LMS:HMDC}}Procedural
resemblances then justify to apply our concepts for alteration
also to real-life QCD: In hadron phenomenology, the behaviour of
the effective threshold may be fixed by fitting predicted
ground-state observables, for instance, hadron masses, to their
experimentally observed values. The decay constant $f_P$ of a
pseudoscalar meson $P\equiv(Q\,\bar q)$ of mass $M_P,$ composed of
a heavy quark $Q$ and a light quark $\bar q,$ is defined by
$(m_Q+m_q)\,\langle0|\bar q\,{\rm
i}\,\gamma_5\,Q|P\rangle=f_P\,M_P^2.$ With
$\overline{m}_c(\overline{m}_c)=(1.279\pm 0.013)\;{\rm GeV}$ for
the $\overline{\rm MS}$ $c$-quark mass, the outcome of our
approach for the decay constants of the charmed $D$ and $D_s$
mesons exhibits a perfect agreement with the findings of lattice
gauge theory and experiment \cite{LMS:HMDC}:
$f_D=\left(206.2\pm7.3_{\rm QCD}\pm5.1_{\rm
syst}\right)\mbox{MeV}$ and $f_{D_s}=\left(245.3\pm15.7_{\rm
QCD}\pm4.5_{\rm syst}\right)\mbox{MeV},$ where the QCD error
comprises all uncertainties of the various (perturbative and
nonperturbative) QCD parameters.

\section{Brief Summary: Observations, Findings, Results,
Conclusions, and Outlook \cite{LMS:QCD,LMS:HMDC}}Clearly, the QCD
sum-rule formalism offers plenty of room for increasing its
predictive power:\begin{itemize}\item The exact effective
continuum thresholds {\em do depend\/} on both Borel parameter and
the relevant momenta; moreover, they are {\em not\/} universal
(that is, they vary with the correlator under study).\item As a
whole, our proposed modifications and improvements raise
dramatically the {\em accuracy\/} of the traditional sum-rule
predictions and yield tenable estimates of their intrinsic
uncertainties.\item The blatant quantitative {\em similarity\/} of
our hadron-parameter extraction {\em procedures\/} in potential
models and in QCD {\em calls for QCD applications\/}, {\em e.g.},
to pseudoscalar-meson form factors \cite{BLM-PSMFFs}.\item Our
sophisticated approach {\em reconciles\/} QCD sum rules and
experiment (as depicted in Fig.~1).\end{itemize}


\begin{figure}[!ht]\begin{center}\begin{tabular}{c}
\includegraphics[width=7.48cm]{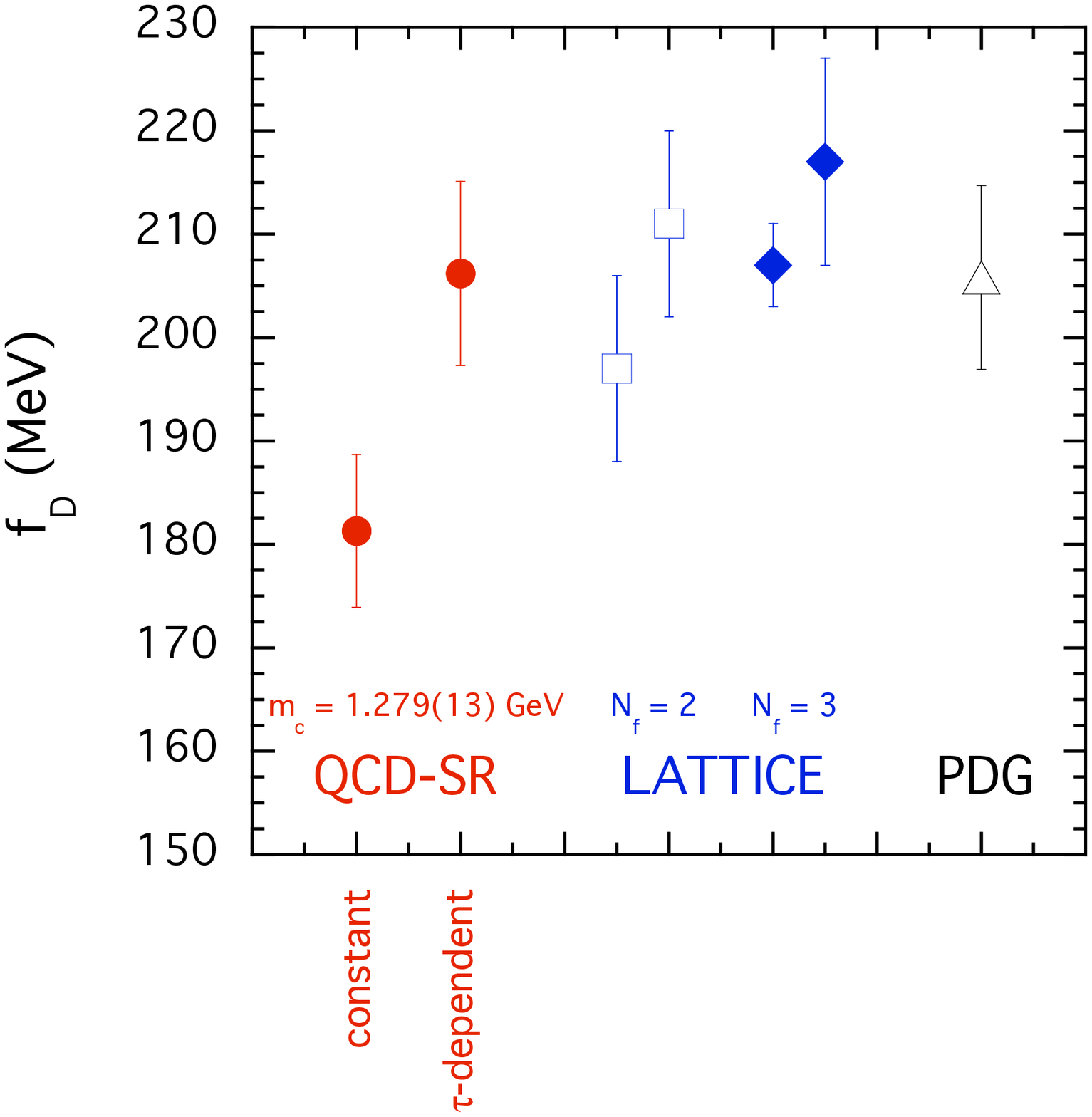}\\[1ex]
\includegraphics[width=7.48cm]{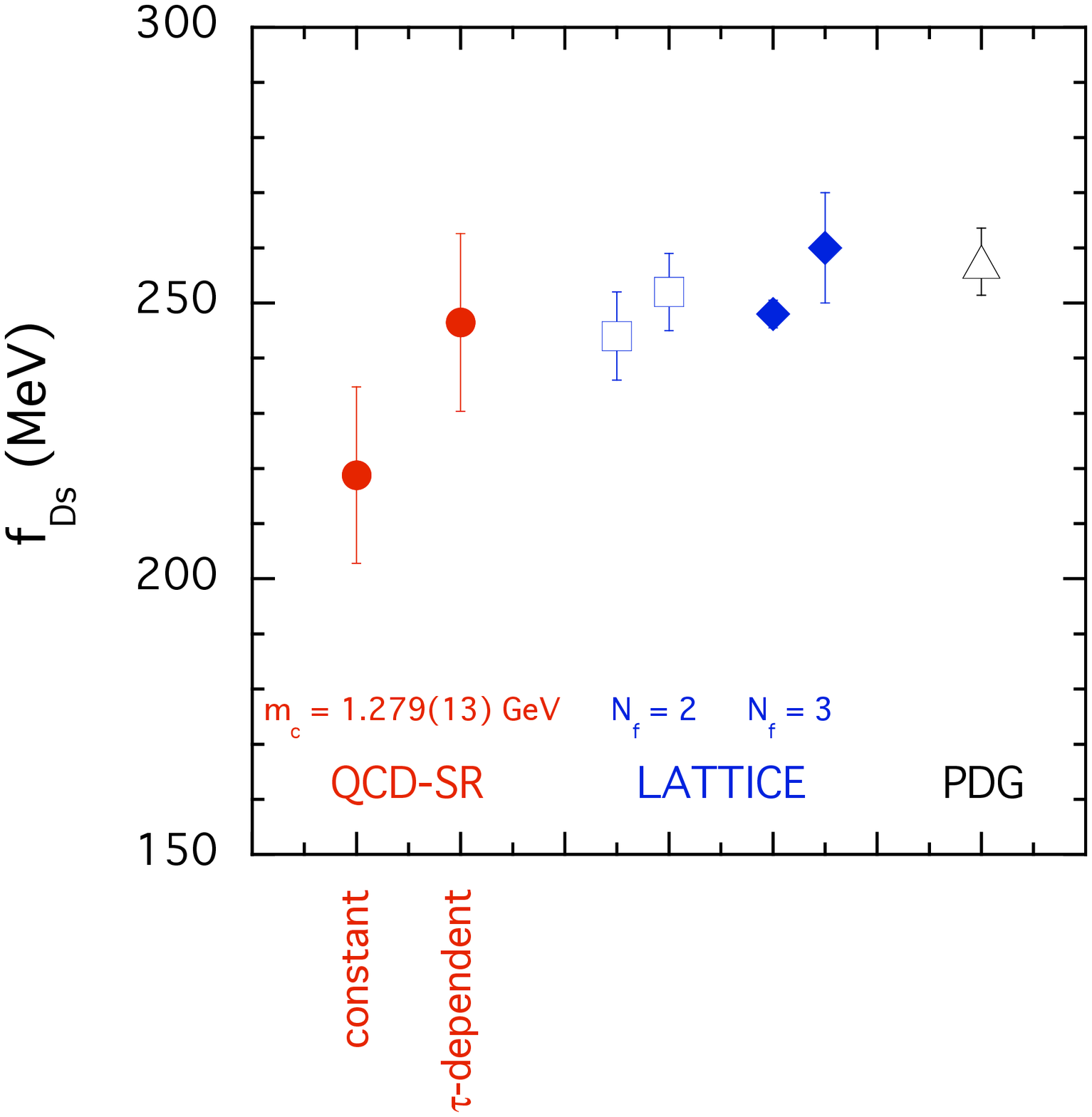}\end{tabular}
\caption{The dependence of the effective thresholds on the Borel
parameter $\tau$ visibly improves the sum-rule results for the
decay constants $f_{D_{(s)}},$ as a comparison with the findings
of lattice QCD and experiment~reveals.}\label{Fig:DCD(s)}
\end{center}\end{figure}

\acknowledgments{D.M.\ is grateful for financial support by the
Austrian Science Fund (FWF), project no.~P22843.}

\end{document}